\documentclass[10pt,conference]{IEEEtran}

\ifCLASSINFOpdf
  \usepackage[pdftex]{graphicx}
  \graphicspath{{./}{../pdf/}{../jpeg/}}
  \DeclareGraphicsExtensions{.pdf,.jpeg,.png}
  \usepackage[caption=false, font=footnotesize]{subfig}
\else
\fi

\usepackage{amsmath}
\usepackage[cmintegrals]{newtxmath}

\usepackage{algorithmicx}
\usepackage{algorithm}
\usepackage{algpseudocode}
\makeatletter
\newcommand{\removelatexerror}{\let\@latex@error\@gobble}
\makeatother

\usepackage{cite}

\usepackage{tabularx}
\usepackage{booktabs}

\usepackage{color,soul}
\setulcolor{red}
\sethlcolor{yellow}
\usepackage{xcolor}

\renewcommand\hl[1]{#1} 
\usepackage{makecell}

\begin{document}
\bstctlcite{IEEEexample:BSTcontrol}

\title{Enabling Resilient and Real-Time Network Operations in Space: A Novel Multi-Layer Satellite Networking Scheme}

\author{\IEEEauthorblockN{Peng Hu}
\IEEEauthorblockA{
\IEEEauthorblockA{\textit{Digital Technologies Research Center, National Research Council of Canada}
}
\IEEEauthorblockA{\textit{1200 Montreal Road, Ottawa, ON K1A 0R6. Canada}
}\\
}
\IEEEauthorblockA{ \vspace{-8pt}
\IEEEauthorblockA{\textit{Faculty of Mathematics, University of Waterloo}
\IEEEauthorblockA{\textit{200 University Ave W., Waterloo, ON N2L 3G1. Canada}}}
}
}

\markboth{Journal of \LaTeX\ Class Files,~Vol.~00, No.~0, 00~2022}%
{Shell \MakeLowercase{\textit{et al.}}: Bare Demo of IEEEtran.cls for IEEE Journals}

\maketitle

\begin{abstract}
Recently advanced low-Earth-orbit (LEO) satellite networks represented by large constellations and advanced payloads provide great promises for enabling high-quality Internet connectivity to any place on Earth. However, the traditional access-based approach to satellite operations cannot meet the pressing requirements of real-time, reliable, and resilient operations for LEO satellites. A new scheme is proposed based on multi-layer satellite networking considering the advanced Ka-band and optical communications payloads on a satellite platform. The proposed scheme can enable efficient and resilient message transmissions for critical telecommand and telemetry missions through different layers of satellite networks, which consist of LEO, medium-Earth-orbit (MEO), and geostationary (GEO) satellites. The proposed scheme is evaluated in a 24-hr satellite mission and shows superior performance improvements compared to the traditional operations approach.
\end{abstract}


%
\IEEEpeerreviewmaketitle

\section{Introduction}
\IEEEPARstart{W}{ith} the increasing launches of non-geostationary (NGSO) satellites, our future Internet infrastructure will be heavily dependent on space-based assets. A variety of critical systems such as mobile communications, transportation, financial services, Earth observation, and defence will rely on satellite-based systems, including non-terrestrial network (NTN) and terrestrial network (TN) segments. Although the upcoming low-Earth orbit (LEO) satellites in large constellations have shed light on an enhanced high-quality Internet, the traditional satellite operational approach has not responded well to the critical challenges imposed by the fast-increasing space assets. A new approach is needed to provide resilient and efficient operations for LEO satellite networks (satnets).

Satellite operations rely on the essential service of transferring control messages to support various Telemetry, Tracking, and Command (TT\&C) missions. Such a service is based on the standards set out by the Consultative Committee for Space Data Systems (CCSDS), where the telemetry (TM) and telecommand (TC) transmission protocols are discussed. For example, the TC packet transmission used in critical spacecraft control is discussed in \cite{CCSDS232.0-B-4}, where the typical approach to TC missions is relatively straightforward: TC packets are transferred from an operations center collocated with a ground station (GS) to the target satellite upon a access opportunity. However, such access opportunities can be meager. For example, based on our extensive access analysis, Starlink's constellation on February 9th, 2022, in a 24-hr mission with a GS located in Iqaluit or Calgary in Canada with 25$^{\circ}$ minimum elevation, the per-satellite GS access opportunity is less than 1.5\%. Even if a global GS network is used, such as the well-known KSAT global GS network with 27 GSs, Starlink's per-satellite GS access opportunity is only increased to around 10\%. The analysis indicates that if a GS attempts to send a TC message to a LEO satellite, there is only $\leq$ 10\% of the time to be successful, where the remaining $\geq$ 90\% of the time is in a waiting status. This waiting time significantly decreases the efficiency of a satnet operation (SatNetOps) and opens the door for possible satellite attacks due to the vulnerabilities of various space-to-ground and inter-satellite links (ISLs) \cite{Giuliari21}. Therefore, there is a clear gap in satellite operations where resilience and efficiency need to be significantly enhanced. The satnets need to be utilized through multi-layer networking (MLN) for the fundamental service in TT\&C-related missions.

The communication and networking between NTN components such as the satnets in LEO, medium-Earth-orbit (MEO), and geostationary (GEO) were not possible due to the limited capability in terms of bandwidth and inconsistent payload used in traditional satellites. This limitation has been relaxed through the recent advancements in satellite communications technologies, where the technologies used on space and ground components are being converged. The intensive use of the Ka/Ku-band payloads on recently launched satellites has been seen. Furthermore, due to the recent development in free-space optical (FSO) communications for satellite communication, such as the promising experimentation of FSO-based space-ground links, and the adoption of FSO from the leading satellite manufacturers/operators, future satellites will be equipped with FSO payloads, significantly improving the bandwidth for space-to-ground and ISLs. Because of the characteristics of the radio-frequency (RF) and optical signals concerning atmospheric effects, RF and FSO payloads are expected to complement each other for future networks such as 6G \cite{Araniti21}.

To respond to the challenges with the consideration of advanced satellite platforms, we propose a novel SatNetOps scheme based on MLN. Compared to the traditional operational approach, the proposed scheme can considerably improve the effectiveness of TC missions in terms of timing performance, resilience, and reliability, and applicable satnets at the LEO/MEO/GEO layers can be utilized for resilience assurance. The results are evaluated extensively with MATLAB, considering the typical commercial satellite constellations. The remainder of the paper is structured as follows. The related work is discussed in Section II. The proposed MLN architecture, system model, and SatNetOps scheme are presented in Section III. The evaluation of the proposed scheme is discussed in Section IV. Section V concludes the paper and states future work.

\section{Related Work}

The revitalized interest in LEO satellite systems comes from the reduced satellite/launch costs, expanded ecosystem, and the capability of bringing low-latency broadband Internet access with global coverage. The idea of satellite networking in different layers can be rooted in multi-layer satellite networking (MLSN) \cite{Nishiyama13}. MLSN was coined as an approach to improving the throughput. The multi-layer networking may occur via multiple shells of a large constellation with LEO satellites, such as Starlink's mega-constellation \cite{Pachler21}, although the cross-shell networking is considered complex \cite{Cakaj21}. Pachler \textit{et al.} \cite{Pachler21} showed the use of optical inter-satellite links (ISLs) on Telesat, Amazon Kuiper, and Starlink constellations can almost double the system throughput of each constellation compared to the non-optical ISLs. A cooperative communication multi-access scheme in MLSN was recently proposed in \cite{Ge2021}. Inmarsat has recently realized a real-world solution for the LEO satellite operations in late 2021 for using the GEO satellites for operations missions of LEO satellites. The SES also implied the benefits of using their upcoming O3b mPOWER fleet for satellite networking. However, the intuitive MLN strategies do not necessarily guarantee the expected performance with the lack of study in algorithm design, use cases, and analysis. 

Current satellite operations are based on individual spacecraft in isolation from other data traffic missions. A ground network operations centre relies on non-real-time operational efforts in the TT\&C processes using standardized communications. CCSDS has standardized the TC space data link protocols (TC-SDLP) for TC missions, where the latest version released in 2021 \cite{CCSDS232.0-B-4} supports the sequence-controlled (Type-A) and expedited (Type-B) services for TC missions with different priorities. Both services have the same frame format, addressing, segmentation and blocking mechanisms. Type-A services support the Automatic Repeat Request (ARQ) for flow control, which is unavailable for Type-B services. Type-B services are used in ``exceptional operational circumstances'' such as spacecraft recovery, or flow control is provided at the upper layers \cite{CCSDS232.0-B-4}. Furthermore, the TC-SDLP-based service only supports unidirectional and asynchronous services with no predefined timing rules specified. With the substantial number of satellites in space, the traditional operations process comes with challenges in efficiency. The timing efficiency for the operations of LEO satellite constellations is also hardly seen in the state-of-the-art.  

The timing performance of the NGSO satellite operational mission depends on the satellite communications systems. Recent systems have broadly adopted the Ka/Ku-band for RF payloads and FSO for optical communication payloads. Since the first test of optical links for space missions in November 2014 \cite{Zech15}, FSO communication is expected to be well adopted by the space industry in the future. The recent FSO-related initiatives from NASA \cite{Park19_NASA}, and commercial LEO satellite constellations further indicate the adoption of FSO-based space-to-space and ground-to-space links. However, the analysis considering both of the payloads for satellite operations is lacking in the literature. 

\section{The Proposed SatNetOps Scheme}

\subsection{An MLN Architecture for SatNetOps}
\begin{figure}[!ht]
\centering
\includegraphics[width=0.95\linewidth]{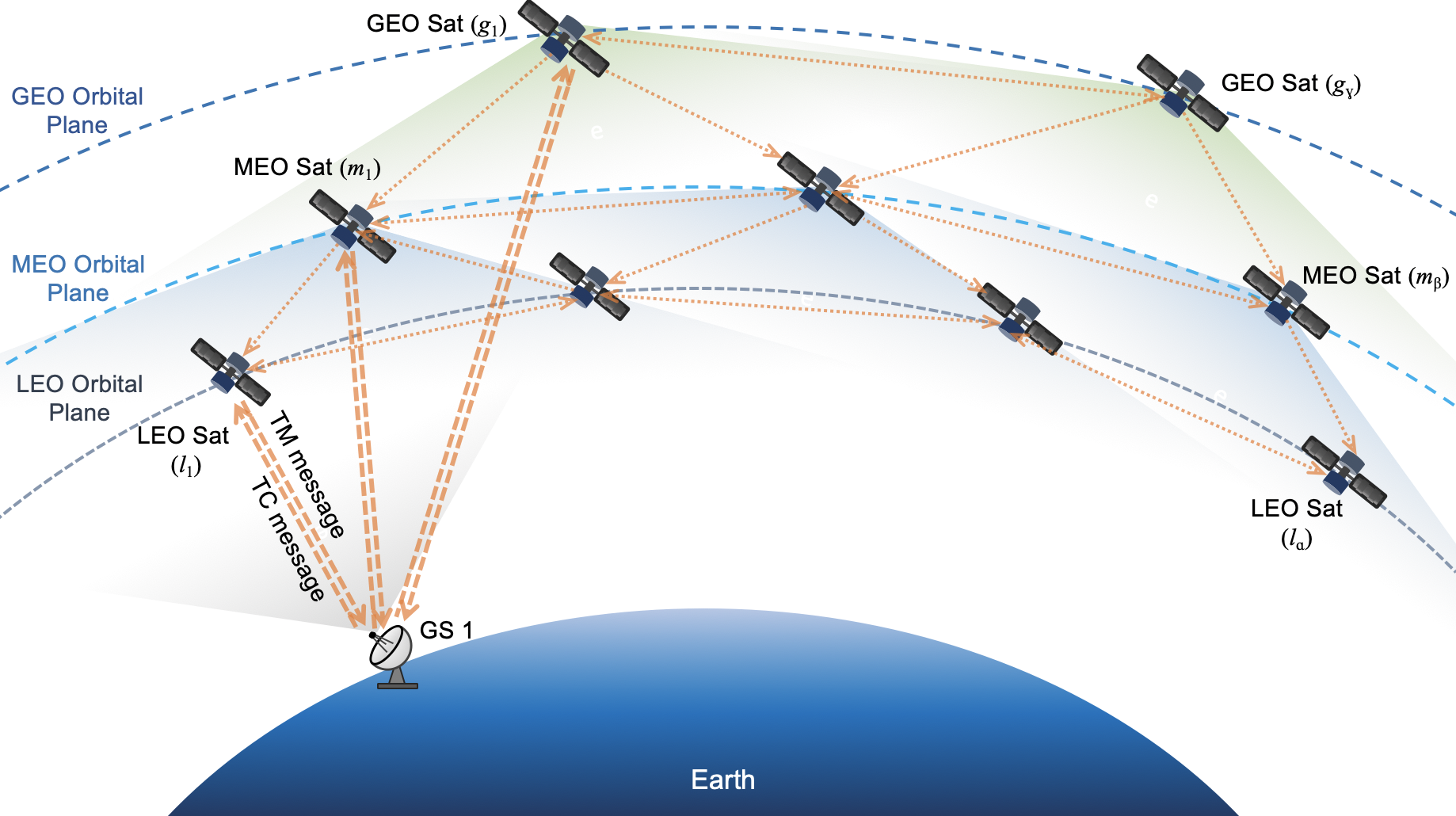}
\caption{Illustration of a multi-layer networking architecture for SatNetOps.}
\label{Fig:illustraion}
\end{figure}
The conceptual layer in our proposed MLN architecture refers to a LEO, MEO, or GEO satellite constellation as a satnet. A 3-layer network is proposed in the MLN architecture as depicted in Fig. \ref{Fig:illustraion}, consisting of GEO, MEO, and LEO satnets. For each satnet, there can be a constellation with one or more shells. However, from the satellite operational perspective, multi-shell communication adds much complexity and is not usually used in satnets, especially for GEO and MEO satellites. Therefore, we do not assume cross-shell communication within a constellation. Specifically, the MLN architecture discussed in this paper considers a basic configuration consisting of one GEO constellation, one MEO constellation, and one LEO constellation, each of which has one orbital shell. The inter- and intra-satellite communication modes between these constellations are summarized as follows: 
\begin{itemize}
    \item A GEO satellite can communicate with its adjacent GEO satellites,
    MEO/LEO satellites, and possible GSs within its coverage. 
    \item MEO satellites can communicate with other MEO satellites, LEO satellites, and GS within their coverage.
    \item LEO satellites can communicate with the inter/intra-LEO satellites and GS within their coverage. 
\end{itemize}

Fig. \ref{Fig:illustraion} shows some example links between the satellites and GS in the proposed MLN architecture, where $l_1$, $m_1$, or $g_1$ represents a source satellite at a satnet layer that can be directly contacted by GS 1 and $l_\alpha$ represents the target LEO satellite, which can be contacted by a destination satellite (e.g., $m_\beta$ or $g_\gamma$) at a satnet layer. Due to the typical satellite network setups and the considerations of size, weight and power (SWaP) on NGSO satellites, we do not consider the communication from LEO/MEO to GEO satellites. 

Based on the proposed MLN architecture, the satnet nodes in LEO, MEO, and GEO constellations are denoted as $L=\{l_1, l_2, ..., l_\alpha\}$, $M=\{m_1, m_2, ..., m_\beta\}$, and $G=\{g_1, g_2, ..., g_\gamma\}$, respectively. The number of satnet nodes in LEO, MEO, and GEO constellations is $n(L)=\alpha$, $n(M)= \beta$, and $n(G)=\gamma$. The goal of the MLN-based SatNetOps scheme is to minimize the latency of a TM/TC message transmission and maintain a high level of resilience and reliability of message transfer.

We can see the combined use of LEO, MEO, and GEO constellations as illustrated in Fig. \ref{Fig:illustraion} would enhance resilience, where a GEO satellite has a high availability due to its fixed location above the Earth's equator, compared to a MEO/LEO satellite. However, a GEO satellite's availability to a GS depends on the GS' elevation and location. If a GS is far beyond the equator region, it may not be well covered by a GEO satellite. Overall, due to the dynamic nature of the satnets, the access opportunities between satellites and GSs are variable and need to be determined through calculations concerning the minimum elevations, constellation and many other configuration parameters of a satnet system (including ground and space components). Also, the distance between adjacent inter-satellites in an orbital plane is fixed, but that may be variable between adjacent intra-satellites. These calculations will be considered in our evaluation in Section IV.  

\subsection{Proposed MLN Scheme}
The proposed MLN scheme for TC missions is described in Alg. \ref{Alg:mln_scheme}, where we can see the scheme ensures the resilience by design in that LEO, MEO, or GEO satnets used for transferring the TC messages are employed when necessary. In Alg. \ref{Alg:mln_scheme}, $GS$ represents the set of GSs where the $j$-th element $GS(j)$ is used. The target LEO satellite is the $i$-th satellite in $L$. The configuration parameters, such as minimum elevations for satellites, are stored in $V$ by satnet layers, which can be retrieved through the layer index $q$. In the procedures $\textsc{SendToLeoSatNet()}$, $\textsc{SendToMeoSatNet()}$, and $\textsc{SendToGeoSatNet()}$ in Alg. \ref{Alg:mln_scheme}, the transmission status is returned and the path of a satnet is determined based on the intra/inter-plane hops and the determination of the nearest satellite in a satnet to the target LEO satellite is included. 

The proposed scheme can utilize LEO, MEO, and GEO to relay packets to the target LEO satellite. Based on Alg. \ref{Alg:mln_scheme}, when a SatNetOps Center attempts to send a TC packet, if there is a LEO satellite that can access the GS, a LEO satnet will be used. If the LEO satellite is inaccessible, then the scheme seeks the MEO satnet. Similarly, if both LEO and MEO satellites are inaccessible, the GEO satnet will be used. The path calculation can occur dynamically due to constant satellite movements. Each LEO/MEO/GEO satellite will determine the best satellite for the next hop based on the slant distance and the minimum elevation. The inter- and intra-plane satellites on the path are calculated based on the well-adopted configuration that each LEO satellite can access up to four neighbouring satellites in the same direction.

\begin{algorithm}
\small
    \caption{Proposed MLN-based SatNetOps Scheme}
    \label{Alg:mln_scheme}
    \begin{algorithmic}[1] 
        \Procedure{ Prepare}{$L$, $M$, $G$, $GS$, $V$, $i$, $j$, $p$}
            \State $l \gets L(i)$, $s \gets GS(j)$
            \State $O \gets \{L, M, G\}$  \Comment{Three-layer satnet elements in $O$}
            \State $q \gets 1$ \Comment{Start search from LEO satnet}
            \For{ $o \gets O(q)$}
            \State $v \gets V(q)$ \Comment{Get config params for a satnet}
            \State $srcSat \gets \Call{GetNearestSatToGs}{s, o, v}$ 
            \If{$srcSat \not= null$}
                \If{$q == 1$} 
                    \If{$srcSat \not= l$}
                        \State $r \gets \Call{SendToLeoSatNet}{p, srcSat, l}$
                    \Else \Comment{The source LEO satellite is the target}
                     \State $\Call{SendToLeoSatNet}{p, l, l}$
                        \State Exit() \Comment{SatNetOps mission is completed}
                    \EndIf
                \EndIf
                \If{$q == 2$} 
                        \State $r \gets \Call{SendToMeoSatNet}{p, srcSat, l}$
                \EndIf
                \If{$q == 3$} 
                        \State $r \gets \Call{SendToGeoSatNet}{p, srcSat, l}$
                \EndIf
                \If{$r == \text{failed}$ }
                    \If{$q < 3$}
                        \State $q \gets q + 1$ \Comment{Move to the next satnet}
                        \State Continue()
                    \EndIf
                \Else
                    \State Exit() \Comment{SatNetOps mission is completed}
                \EndIf
            \Else
               \State $q \gets q + 1$ \Comment{Move to the next satnet}
               \State Continue() 
            \EndIf

            \EndFor

        \EndProcedure

        \Procedure{GetNearestSatToGs}{$s$, $o$, $v$} 
                \State Calculate the nearest satellite object over $s$ on $o$
				\State \textbf{return} $o$ 
		\EndProcedure
        \Procedure{SendToLeoSatNet}{$p$, $srcSat$, $o$} 
				\State Send packet $p$ from $srcSat$ of the LEO satnet to target LEO sat $o$
		\EndProcedure
        \Procedure{SendToMeoSatNet}{$p$, $srcSat$, $o$} 
				\State Send packet $p$ from $srcSat$ of the MEO satnet to target LEO sat $o$
		\EndProcedure
        \Procedure{SendToGeoSatNet}{$p$, $srcSat$, $o$} 
				\State Send packet $p$ from $srcSat$ of the GEO satnet to target LEO sat $o$
		\EndProcedure
    \end{algorithmic}
\end{algorithm}

The proposed scheme removes the need for direct access opportunities between a GS and a target satellite, which improves timing performance. To quantify the timing performance, let us denote the overall latency of a TC transfer mission by $D$, which consists of four components, i.e., propagation delay, transmission delay, processing delay, and queuing delay. Due to the small size of a TC packet, the average queuing delay and average processing delay per satellite can be assumed to be small constants, $m$ and $k$, respectively. We let $M$ and $c$ denote the size of a TC packet and the speed of light, respectively. The data rates via an RF link using Ka and L bands and an FSO link are denoted by $r_{k}$, $r_{l}$, and $r_{o}$, respectively.

The path length from a GS to the target LEO satellite needs to be considered as it relates to the latency components. For instance, we suppose the distance from a GS to the source LEO satellite, source MEO satellite, and source GEO satellite is $d_0$, $d_1$, and $d_2$, respectively; the distance from a source MEO satellite to a destination MEO satellite is $d_3$; the distance from a destination MEO satellite to a target LEO satellite is $d_4$; the distance from a source GEO satellite to a destination GEO satellite is $d_5$; the distance from the destination GEO satellite to a target LEO satellite is $d_6$; and the distance from the source LEO satellite to the target LEO satellite is $d_7$. Let the total path length of these hops be $L_{h}$. In the proposed scheme, $L_h = d_0 + d_7$ if a LEO satnet is used, $L_h = d_1 + d_3 + d_4$ if a MEO satnet is used, or $L_h = d_2 + d_5 + d_6$ if a GEO satnet is used. In this sense, the latency can be derived in (\ref{Eq:latency}) as:

\begin{equation}\label{Eq:latency}
 D =  \sum^{n_h}_{i=1}{\left( \frac{d(i)}{c} + \frac{M}{r(i)} + m + k \right)},
\end{equation}
where $n_h$ is the hop count, and $d(i)$ and $r(i)$ are the length and data rate of the $i$-th hop.

If we take multiple hops into consideration, the overall reliability $\Phi$ is defined in (\ref{Eq:reliability}) as:
\begin{equation} \label{Eq:reliability}
    \Phi = \left( 1 - \prod_{i=1}^{n_h}{(1-\phi(i))} \right),
\end{equation}
where $\phi(i)$ is the reliability of the $i$-th hop on a path. 

By considering the link reliability (e.g., possible propagation impairments, etc.), satellite lifetime, and link availability, we adopt the result \cite{Emily2018} that the reliability of a GEO satellite link is higher than that of an NGSO link, i.e., $\phi(i)_{GEO} \geq \phi(i)_{MEO} > \phi(i)_{LEO}$.

The resilience measure is defined in (\ref{Eq:resilience}) as:
\begin{equation} \label{Eq:resilience}
    \lambda = \frac{K_0}{K},
\end{equation}
where $K$ is the number of total runs of TC missions and $K_0$ is the number of successful runs of TC missions.

\begin{table}[ht]
\centering
\renewcommand{\arraystretch}{0.95}
\setlength{\tabcolsep}{1.2pt}
\caption{Scenarios (S1-S4) for Evaluation Based On Link Types}
\label{Tbl:eval_scenarios}
\begin{tabular}{c c c c c}
\toprule
{} &
{\bfseries GS-LEO/MEO/GEO} & {\bfseries LEO/MEO-LEO} & {\bfseries GEO-GEO/LEO} & {\bfseries MEO-MEO} \\ 
\midrule
{\bfseries S1} & Ka & Ka & Ka & Ka\\
{\bfseries S2} & Ka & Ka & L & Ka \\
{\bfseries S3} & FSO & Ka & Ka & Ka \\
{\bfseries S4} & FSO & FSO & FSO & FSO\\
\bottomrule
\end{tabular}
\end{table} 

\begin{table}[ht]
\centering
\setlength{\tabcolsep}{0.5em} 
\renewcommand{\arraystretch}{0.95}
\setlength{\tabcolsep}{2pt}
\caption{Key Simulation Parameters}
\label{Tbl:parameters}
\begin{tabular}{c r l}
\toprule
{\bfseries Parameter} & {\bfseries Value} & {\bfseries Notes}\\ 
\midrule
$(lat, lng)$ & (51.0447, -114.0719) & GS (SatNetOps center)  coordinate\\
$minElev1$ & 25$^\circ$ & Min. elevation of the GS\\
$minElev2$ & 10$^\circ$ & GEO/MEO sats to GS elevation\\
$\{\alpha, \beta, \gamma\}$ & \{78, 20, 3\} & Satnet numbers\\
$r_{o,gl}$ & 1.8 Gbps \cite{Zech15} & Rate of the FSO GEO-LEO link \\
$r_{r,gl}$ & 324 Mbps \cite{Emily2018} & Rate of the RF GEO-LEO link \\
$r_{k}$ & 324 Mbps & Rate of the Ka-band link \\
$r_{l}$ & 150 kbps & Rate of of the L-band link \\
$M$ & 1024 B & TF size of a TC packet\\
$T$ & 24 hr & Mission duration \\
$T_{start}$ & 2022-08-14 01:00:00 & Satellite scenario start datetime \\
$T_{end}$ & 2022-08-15 01:00:00 & Satellite scenario end datetime \\
$T_{sample}$ & \hl{3600 s} & Sample time \\
$k$ & 100 $\mu$s & Avg. processing delay on a satellite\\
$m$ & 100 $\mu$s & Avg. queuing delay on a satellite\\
$\phi_1(i)$ & 0.9980 & Reliability of a LEO ISL\\
$\phi_2(i)$ & 0.9990 & Reliability of a GEO/MEO ISL\\
$\phi_3(i)$ & 0.9990 & Reliability of a GEO-LEO link\\
$\phi_4(i)$ & 0.9985 & Reliability of a MEO-LEO link\\

\bottomrule
\end{tabular}
\end{table}

\section{Performance Evaluation}
\begin{figure}[!t]
\centering
\includegraphics[width=0.95\linewidth]{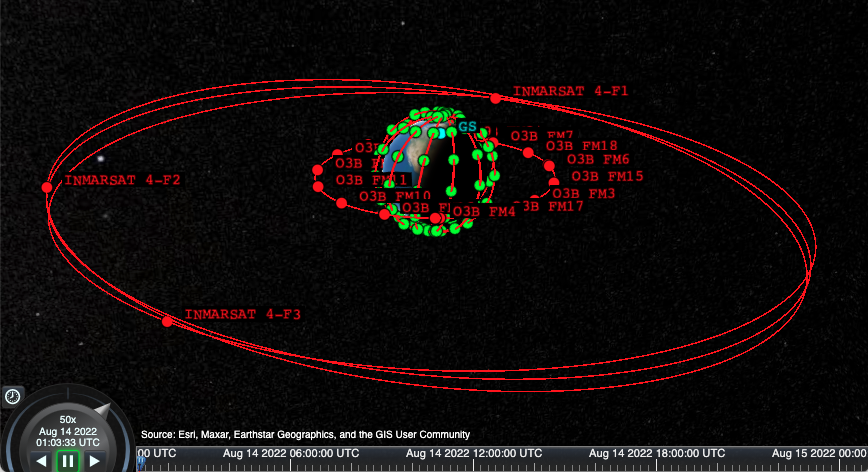}
\caption{The satellite scenario in MATLAB}
\label{Fig:scenario_view}
\vspace{-15pt}
\end{figure}

The proposed MLN-based SatNetOps scheme is evaluated using the MATLAB Satellite Communications Toolbox in simulation programs. The satellite scenario and parameters are shown in Fig. \ref{Fig:scenario_view}, Table \ref{Tbl:eval_scenarios}, and Table \ref{Tbl:parameters}. In the satellite scenario, the existing Inmarsat-4 GEO and SES O3b MEO satellites are used, where the public two-line element (TLE) data for these satellites is used in the simulation. The Telesat LEO satellites in polar orbit are used, where the ephemeris is propagated based on the public filing data, where the inclination is 98.98$^{\circ}$ and altitude is 1015 km with 78 satellites in 6 orbital planes. The Telesat polar constellation can well cover the northern regions and have support for optical communication payloads. The GS is located in Calgary, Alberta, a city in western Canada, with a minimum elevation angle of 25$^{\circ}$. To obtain results that can cover general payload options, we consider both RF and FSO connectivity options that mainstream communications payloads can realize on space-to-space and ground-to-space links in four typical scenarios S1-S4. Table \ref{Tbl:eval_scenarios} shows the combinations of the Ka/L-band and FSO in S1-S4, where S1 and S2 are RF bands and S3 and S4 include the FSO links. In S2, the L-band is considered for GEO-GEO/LEO links for compatibility reasons. The L-band has been widely available on legacy GEO/LEO satellites. S3 and S4 reflect the combination of RF and FSO links for future NTN communications indicated in \cite{Araniti21}.

The key simulation parameters are shown in Table \ref{Tbl:parameters}, where the data rates for \hl{L-band} and Ka-band links are based on the values in \cite{Zech15, Emily2018}. The proposed MLN scheme for SatNetOps is compared with two typical schemes: one is the traditional operational scheme (referred to as ``traditional''), where TC packets are sent once the target satellite has access to the GS; and the other is the GEO-based scheme (referred to as ``GEO Only''), where only GEO satnet is used. From the CCSDS standards for space packet protocol (SPP) and TC-SDLP \cite{CCSDS133.0-B-2, CCSDS232.0-B-4}, a transfer frame (TF) in TC-SDLP has a maximum length of \hl{1024 B}, so we let $M=1024$ B in the evaluation. For the processing delay $D_{pc}$ per satellite, based on the results in \cite{Carlsson2004DelayPI}, indicating the mean delay for UDP/ICMP payload size from 32 B to 1450 B (which applies to $M$) to be  around 100 $\mu$s, we let $k=100~\mu$s. For $m$, we consider a slight latency in the same scale of $k$, where in this case $m=100~\mu$s.

In the LEO constellation, every fourth satellite is picked as a target, ensuring that each orbital plane has a good number of target satellites. The simulations are executed over a 24-hr mission to acquire adequate data, and the results are statistically analyzed based on all iterations. 

\subsection{Simulation Results}
Let us see the performance of the proposed MLN scheme, where the hop count and path length will be examined first as they are related to MLN of the proposed scheme, followed by the latency, resilience, and reliability based on the measures discussed in Section III.

\begin{figure}[!t]
\centering
  \subfloat[\label{Fig:hop_a}]{%
       \includegraphics[width=0.86\linewidth]{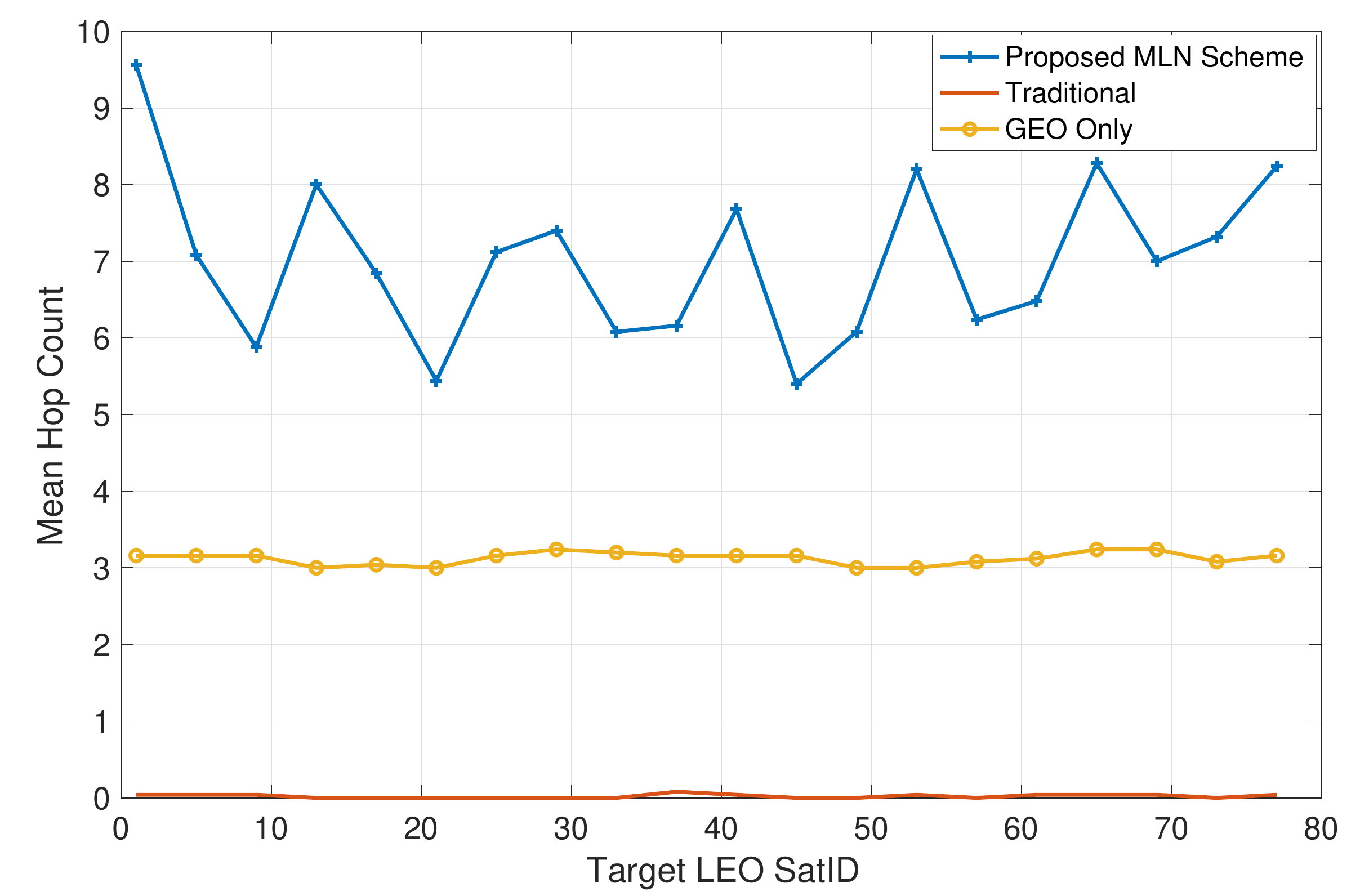}}
    \vfill \vspace{-10pt}
  \subfloat[\label{Fig:hop_b}]{%
        \includegraphics[width=0.86\linewidth]{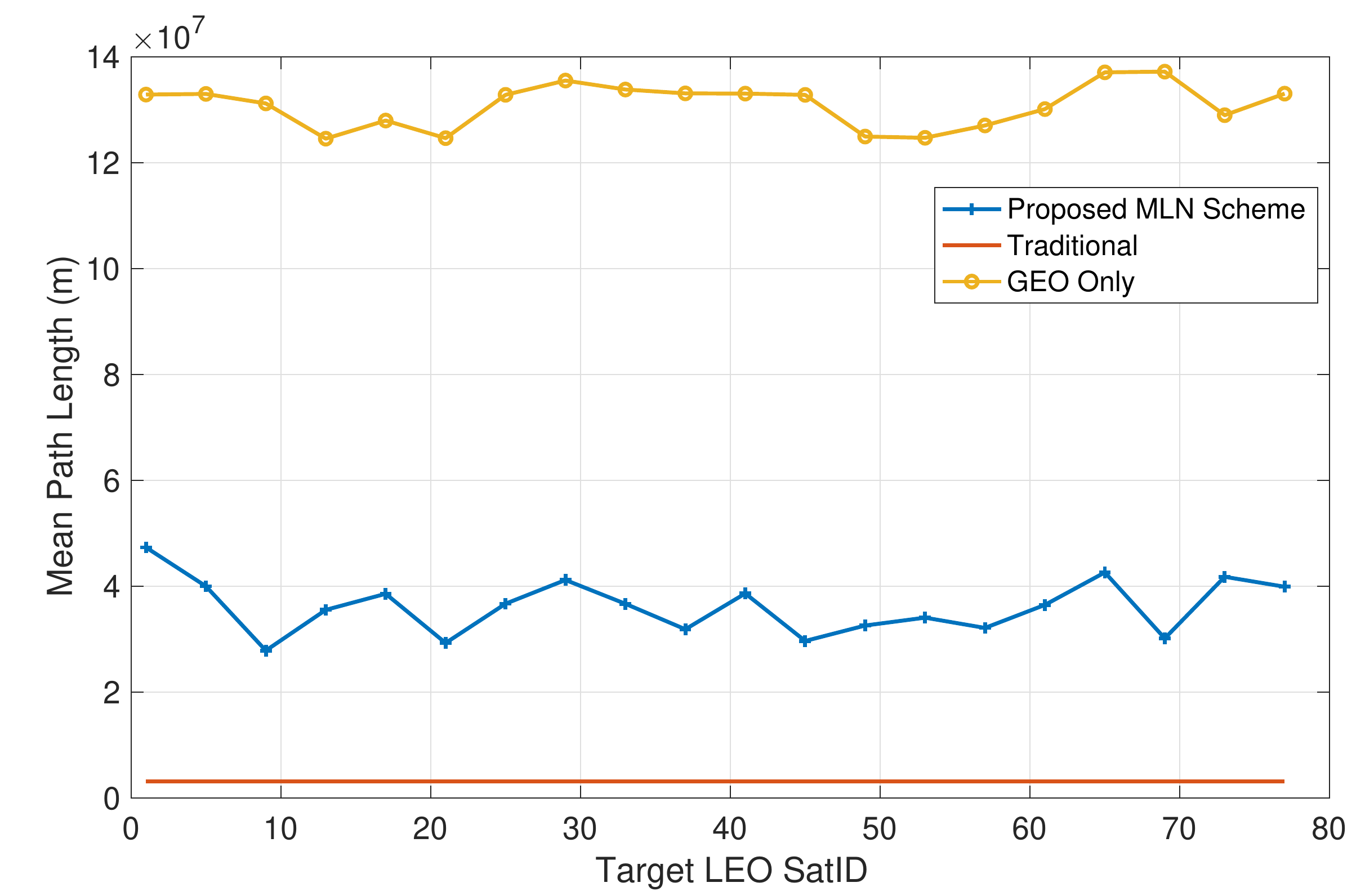}}
\caption{(a) Mean hop count and (b) path length in S1-S4}
\vspace{-15pt}
\label{Fig:hop}
\end{figure}

Fig. \ref{Fig:hop} shows the mean hop count and path length over target LEO satellites, where satellites are indexed by IDs. Because hop counts are invariable to the data rates, the performance does not vary from S1 to S4. From Fig. \ref{Fig:hop_a}, we can see that the proposed scheme uses multiple hops to transfer the TC packet, which meets our expectations well. The ``GEO Only'' scheme involves less and steadier hop counts for all target satellites because of the constant availability of the GEO satellites used in our simulations. In the traditional scheme, a GS sends a TC packet to the target LEO satellite only if it has direct access, so it has the least hop count overall. The mean hop count for the proposed scheme and the ``GEO Only'' scheme is 7.024 and 3.126, respectively. However, the mean path length for the proposed scheme, as shown in Fig. \ref{Fig:hop_b} is 3.6118e+07 m, which is 3.6242 times shorter than that for the ``GEO only'' scheme. The traditional scheme has a mean path length of 3.0633e+06 m due to the low altitude of a LEO satellite in the range of a GS. 

\begin{figure}[!t]
\centering
\includegraphics[width=0.9\linewidth]{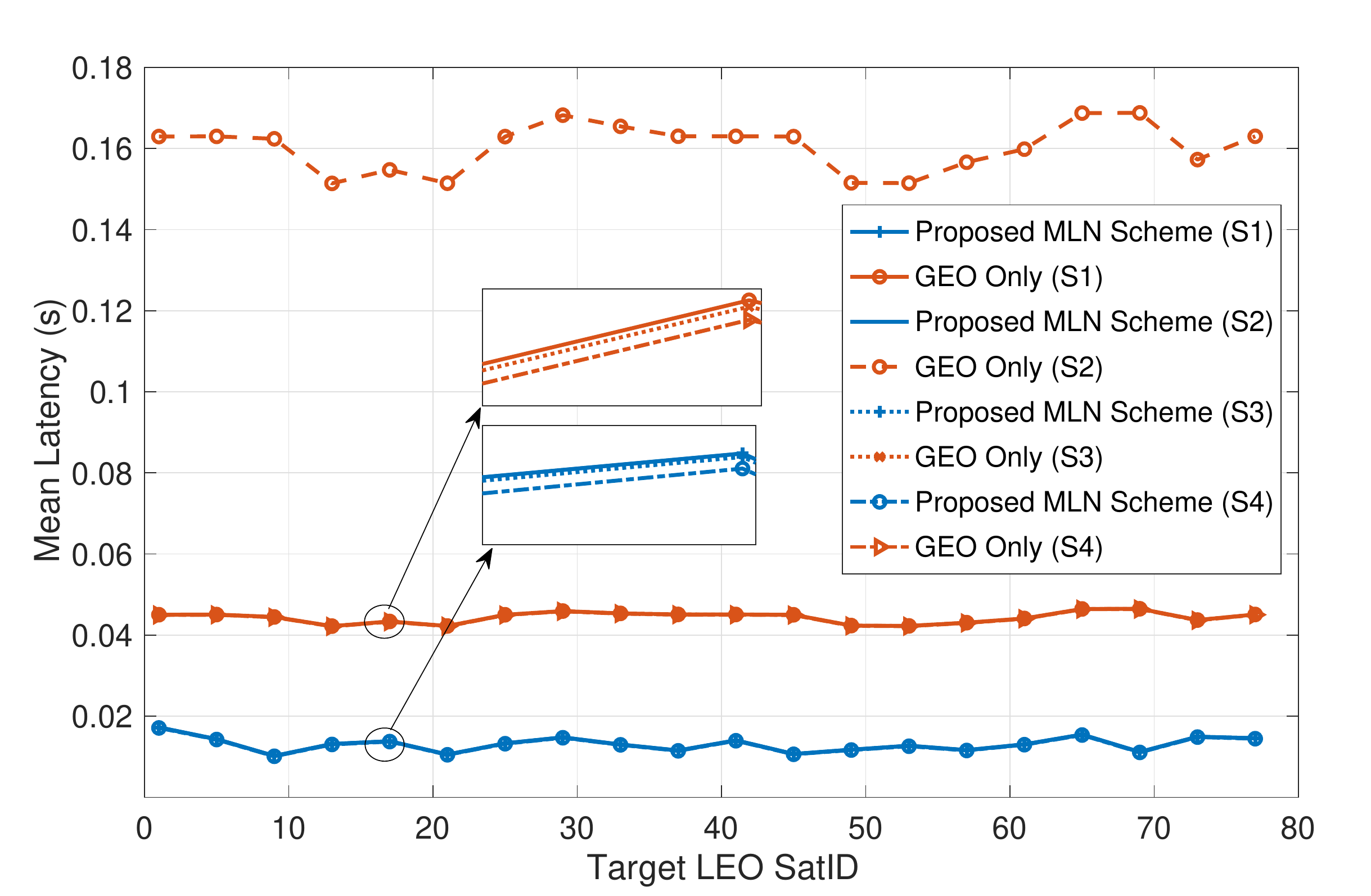}
    \vfill \vspace{-12pt}
\caption{Latency performance in S1-S4}
\label{Fig:latency}
\vspace{-15pt}
\end{figure}

Fig. \ref{Fig:latency} shows the latency performance of the proposed scheme and the ``GEO Only'' scheme. The proposed scheme outperforms the ``GEO Only'' in all scenarios. The overall mean latency of the proposed scheme is 13.1 ms in S1-S3 and 13.0 ms in S4, which outperforms the ``GEO Only'' scheme by 3.3893, 12.2443, 3.3817, and 3.4077 times in S1-S4, respectively. Because of the small size of the TC packet and insignificant path length difference for RF and FSO signal propagation, as shown in Fig. \ref{Fig:latency}, the difference in latency performance in S1-S4 is small (at the scale of milliseconds) in the proposed scheme. 

\begin{figure}[!t]
\centering
\includegraphics[width=0.9\linewidth]{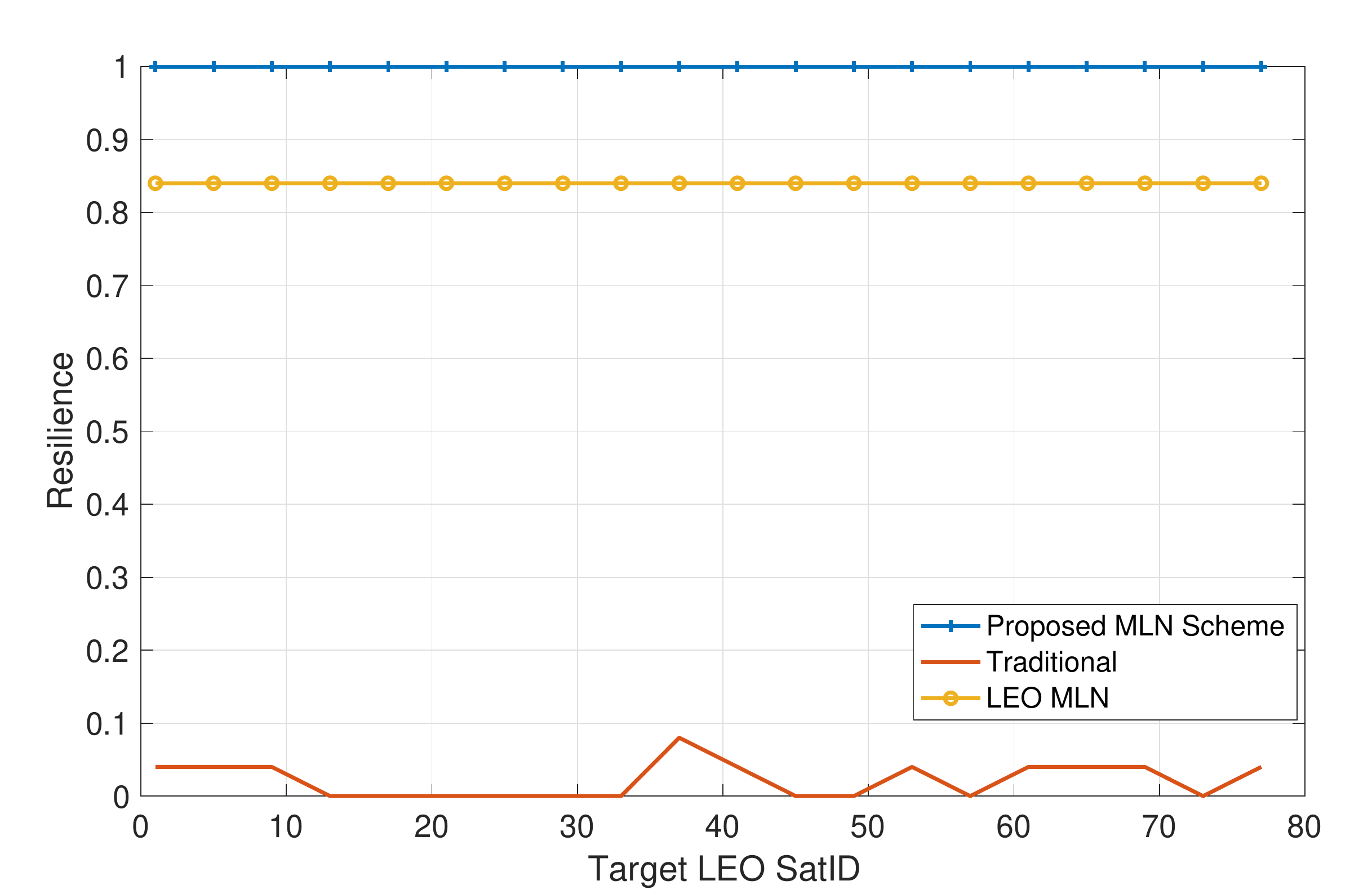}
\caption{Resilience performance}
\vspace{-12pt}
\label{Fig:resilience}
\end{figure}

Fig. \ref{Fig:resilience} shows the resilience performance of the proposed scheme, where the higher the value, the better. Compared to the traditional scheme, which stays at a low resilience below 8\% at all times, the proposed scheme reaches 100\% resilience. To provide another good reference scenario, i.e., when no MEO/GEO satnet layer is available, a modified version of the proposed scheme referred to as ``LEO MLN'' is used. The performance of the ``LEO MLN'' scheme is shown in Fig. \ref{Fig:resilience}, where its resilience is reduced to 84\%. In the satellite mission under evaluation, we found that no GEO satnet is employed because there is always a source MEO satellite above the GS when a source LEO satellite is unavailable. However, even if both source MEO and LEO satellites are unavailable above a GS, our proposed scheme can still reach 100\% resilience as it can automatically use a GEO satnet. It is worth noting that, when the ``GEO Only'' scheme is used, its resilience performance can be 100\% when GEO satnet is available or 0\% when a single point of failure occurs on a link or GEO satellite node. This may introduce a considerable risk to SatNetOps, in addition to its worst timing performance.

\begin{figure}[!t]
\centering
\includegraphics[width=0.9\linewidth]{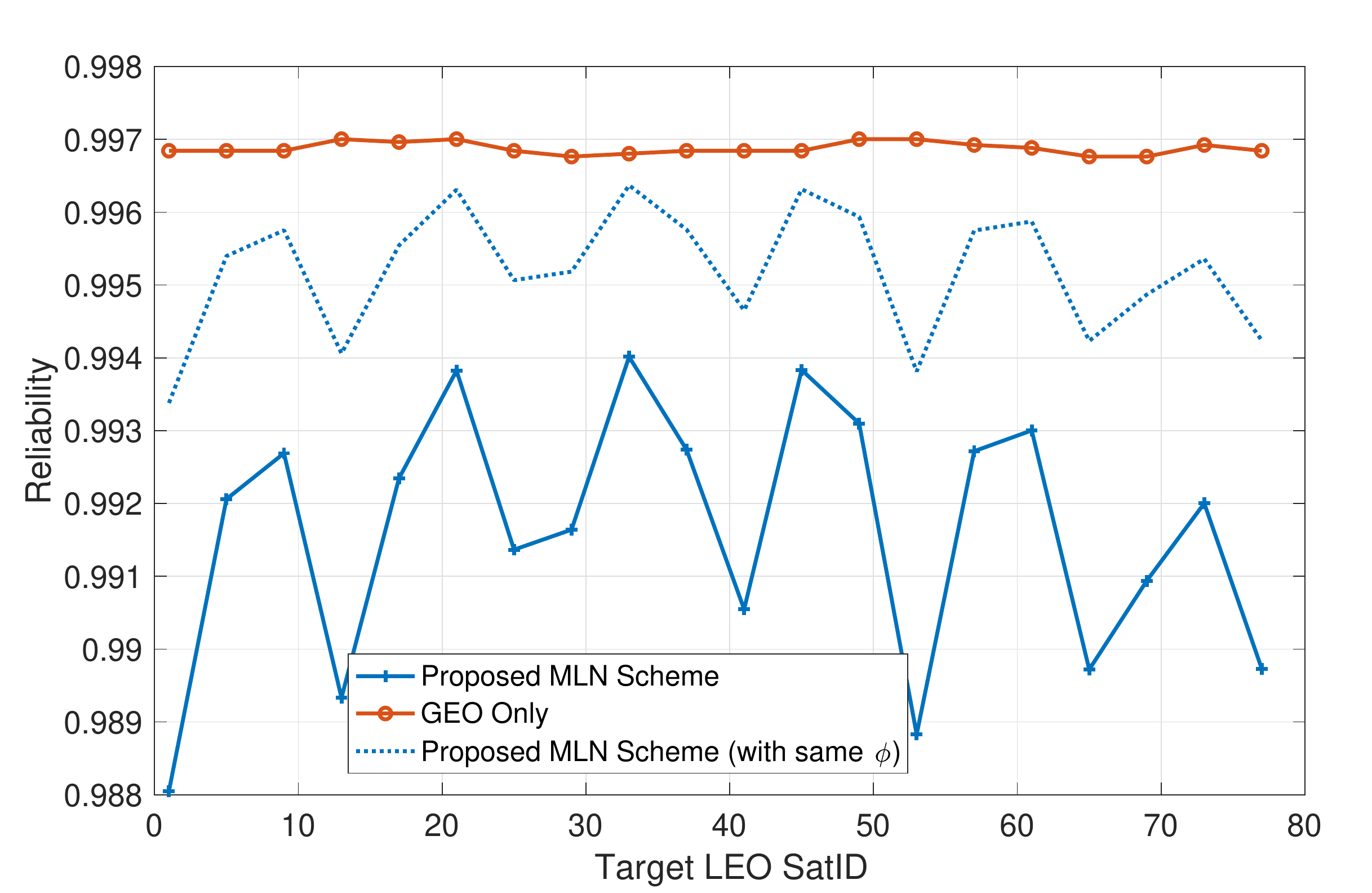}
\caption{Reliability performance}
\vspace{-15pt}
\label{Fig:reliability}
\end{figure}

Fig. \ref{Fig:reliability} shows the reliability between the proposed scheme and the ``GEO Only'' scheme, where we can see that the proposed scheme maintains high reliability, comparable to the ``GEO Only'' scheme. Specifically, the overall mean reliability of the proposed scheme is 99.16\%, compared to 99.69\% for the ``GEO Ony'' scheme. In this sense, the slightly reduced reliability by 0.53\% is considered the cost of the significant timing performance improvements that the proposed MLN scheme brings. In consideration of the possible advancement in the link reliability in the case that all links have the same $\phi$ of 0.999, we can see the reliability of the proposed scheme is improved as shown in the dotted line in Fig. \ref{Fig:reliability}, where the overall mean reliability is 99.52\%.

\section{Conclusion}
The proposed MLN-based SatNetOps scheme can resolve the operational challenges imposed by the increasing number of LEO satellites. With the promising performance in terms of latency, resilience, and reliability, the proposed scheme can benefit the TT\&C missions for real-time or near-real-time operational needs, improve the efficiency of satnet resources, and help reduce the reliance on GSs, which ultimately reduces operating costs in many ways. Additional configuration setups and methods for improving responsiveness based on this paper will be investigated in future work.


%

\section*{Acknowledgment}
This work was supported by the High-Throughput and Secure Networks Challenge program of National Research Council Canada. We also acknowledge the support of the Natural Sciences and Engineering Research Council of Canada (NSERC), [funding reference number RGPIN-2022-03364].


\ifCLASSOPTIONcaptionsoff
  \newpage
\fi



\bibliographystyle{IEEEtran}
\bibliography{./references}
\end{document}